\documentclass[preprint2]{aastex}

\slugcomment{}

\shorttitle{MMSN and migration}

\shortauthors{Crida A.}

\begin{document}

\title{Minimum Mass Solar Nebul{\ae} and Planetary Migration}

\author{Aur\'elien Crida}
\affil{Institut f\"ur Astronomie \& Astrophysik, University of T\"ubingen,\\ Auf der Morgenstelle 10, D-72076 T\"UBINGEN, GERMANY}
\email{crida@tat.physik.uni-tuebingen.de}


\begin{abstract}
The Minimum Mass Solar Nebula (MMSN) is a protoplanetary disk that
contains the minimum amount of solids necessary to build the planets
of the Solar System. Assuming that the giant planets formed in the
compact configuration they have at the beginning of the ``Nice
model'', \citet{Desch2007} built a new MMSN. He finds a decretion
disk, about ten times denser than the well-known Hayashi MMSN. The
disk profile is almost stationary for about ten million years.
However, a planet in a protoplanetary disk migrates. In a massive,
long-lived disk, this question has to be addressed. With numerical
simulations, we show that the four giant planets of the Solar System
could not survive in this disk. In particular, Jupiter enters the
type~III, runaway regime, and falls into the Sun like a stone. Known
planet-planet interaction mechanisms to prevent migration, fail in
this nebula, in contrast to the Hayashi MMSN. Planetary migration
constrains the construction of a MMSN. We show how this should be done
self-consistently.
\end{abstract}



\section{Introduction}

The Minimum Mass Solar Nebula (hereafter MMSN) is the protoplanetary
disk of solar composition that has the amount of metals necessary to
build the eight planets of the Solar System (and the asteroid
belts). From the masses and compositions of the planets, a density of
solids is derived at several locations of the disk. Then, the solar
composition is restored by adding gas, and a smooth protoplanetary
disk density profile is derived. For a review of past works on the
MMSN concept, see the introduction of \citet{Desch2007}. The most
famous version of the MMSN was provided by {\citet{Weiden77} and}
\citet{Hayashi1981}. Assuming that the giant planets have a rocky-icy
core of about 15 Earth masses, and that Jupiter accreted all the solid
material available between 1.55 and 7~AU, {Hayashi} finds a
surface density profile of $\Sigma(r)=1700 (r/1\,{\rm AU})^{-3/2}\
{\rm g.cm}^{-2}$, {similar to the one given by
\citet{Weiden77}}. Of course, the density profile obtained depends
crucially on the position of the planets. Hayashi assumed that the
planets formed where they presently orbit.

A recent model explains several features of the Solar System (the Late
Heavy Bombardment, the orbital distribution of the Trojans of Jupiter,
the orbital elements of the giant planets\ldots) thanks to a late
global instability in the outer Solar System dynamics
\citep{LHB-Gomes,LHB-Morby,LHB-Tsiganis}. In this so-called ``Nice
model'', the four giant planets were in a compact configuration just
after the solar nebula dissipation. Therefore, if one assumes that
this model is true, the {\citet{Weiden77} and} \citet{Hayashi1981}
nebula is out of date. In a recent, very nice article,
\citet{Desch2007} constructed a new MMSN, assuming that the planets
were formed in the disk at their starting position in the Nice
model. He finds that the density distribution required to form the
giant planets in this configuration, including the presence of an
exterior disk of planetesimals, can be very well fitted by a steep
power law density profile. In addition, he studies the time evolution
of this disk. He finds a solution of a decretion disk, that can
survive with an almost unchanged density profile in the giant planets
region for a few million years. This enables the solid cores of the
four giants to reach their isolation masses, and then to slowly
accrete their gaseous envelope. Consequently, this nebula is self
consistent from the planetary formation point of view. The nice
agreement between the initial position of the planets in the Nice
model and a long-lived, almost power-law protoplanetary disk can be
seen as a new plus point for the Nice model.

However, planets in gaseous disks are subject to planetary migration
\citep[see][for a review]{PapaTerq}. How to prevent Jupiter (and
Saturn) from becoming a hot Jupiter (or Saturn) is a longstanding
problem in Solar System formation, but some solutions have been
proposed. \citet{MS01} have shown that a mean motion resonance between
Jupiter and Saturn can enable them to resist the driving inwards by
the disk, and to migrate together outwards. \citet{MC07} found disk
parameters for which Jupiter and Saturn enter the 3:2 mean motion
resonance and then have a negligible
migration. \citet{Morby-etal-2007} then showed that Uranus and Neptune
can also be saved in that case, by capture in resonance with
Saturn. The four giant planets are in the end of the gas disk phase in
a compact, fully resonant configuration. Two of these possible final
configurations can lead to a late global instability of the dynamics,
as required in the Nice model. These works were based on a disk in
which the density slope was low and the gas density at the location of
Jupiter was of the order of the one in the \citet{Hayashi1981} nebula.

In this paper, we address the question of the planetary migration in
the \citet{Desch2007} nebula. In section~\ref{sec:disk}, we describe
our code, that enables a simulation of the entire disk, and of its
global evolution. We also review the disk properties, and show how
they should affect the planetary migration. The results are presented
in section~\ref{sec:isoth} for a locally isothermal disk, and in
section~\ref{sec:energ} for a disk where the energy equation is taken
into account. In every case, the planets are lost into the Sun. In
section~\ref{sec:MMR}, we try to reproduce the \citet{MC07} result in
the \citet{Desch2007} disk, or in a colder disk with same density
profile. {In section~\ref{sub:UN}, we focus on original results on
Uranus and Neptune in some simulations}. For comparison, the migration
in the \citet{Hayashi1981} MMSN in presented in
section~\ref{sec:Hayashi}. {Some numerical issues are discussed in
section~\ref{sec:discussion}. Finally,} our conclusions are presented
in section~\ref{sec:conclu}, and some prospectives are {
suggested}. We show how planetary migration should be considered in
the construction of a Minimum Mass Solar Nebula, and we draw some
ideas towards a MMSN compatible with both planetary migration and the
Nice model.

\section{Numerical settings, disk properties and expected migration}
\label{sec:disk}

\subsection{Code description}

We use the code FARGO \citep{FARGO1,FARGO2}, in its 2D1D
version\footnote{Now publicly available on
\url{http://fargo.in2p3.fr}} \citep{Crida-etal-2007}. In this
version, the classical 2D polar grid of the hydro-code (in which the
planet-disk interactions are computed) is surrounded by a 1D grid. In
the 1D grid, the disk is considered as axisymmetric, and only the
radial component of the equations is computed\,; this is
computationally very cheap, therefore the 1D grid extension can be
very large. The two grids are smoothly connected, so that the disk
evolution is computed accurately over all the 1D grid extension\,: the
evolution of the inner and outer parts of the disk computed in the 1D
grid, influence the disk evolution in the 2D grid\,; and reciprocally
the planet-disk interactions computed in the 2D grid perturb the
global disk evolution.

This property of the code makes it the perfect tool to study the
behavior of the giant planets in the MMSN. Indeed, giant planets
generally open gaps in the disk, and then are locked in the global
viscous evolution of the disk (type~II migration). Most often, this
drives the planets inwards. But in the Desch nebula, the disk is
viscously spreading. Consequently, its global evolution should drive
planets in type~II migration outwards. However,
\citet{Crida-etal-2007} and \citet{Crida-Morby-2007} found that things
are not that simple. If the gap is not completely empty, a corotation
torque is exerted on the planet, gas can pass from the inner to the
outer disk (or reciprocally), and the planet can decouple from the
disk evolution. Type~II migration can then proceed inwards in a
decretion disk, or outwards in a viscous accreting disk. This can be
seen only with the use of the 2D1D algorithm, that permits both an
accurate 2D computation of the planet-disk interaction and the 1D
computation of the global disk evolution. The global disk evolution
also matters for the type~III regime.

The outer edge of the 1D grid is located at $61$~AU from the star, as
prescribed by \citet{Desch2007}. The inner edge is located arbitrarily at
$0.2$~AU. The inner and outer edges of the 1D grid are open, allowing
gas to flow out of the grid.

The rings of the grid are logarithmically distributed\,: $\delta r/r$
is constant through the two grids. The cells of the 2D grid are
squared\,: $\delta r/r=\delta \theta$.

The 2D grid extends from $1$ or $1.6$ to $23.5$~AU, covering the planets
region.

In the computation of the force of the disk on the planets, a part of
their Hill sphere is excluded. To perform this, we calculate in the
code the vector force $\vec{F}_p$ using the following
expression\,:
\begin{equation}
\vec{F}_p = \sum_{\rm cells}
\frac{GM_p\Sigma\,{\rm d}{\cal S}}{s'^3}f(s) \vec{s}\ ,
\label{eq:Force}
\end{equation}
where $G$ is the gravitational constant, $M_p$ the mass of the planet,
${\rm d}{\cal S}$ is the surface of the considered cell, $\vec{s}$ is
the vector from the planet to the center of the cell, $s$ is its
length, and $s'=\sqrt{s^2+\epsilon^2}$ is the smoothed distance to the
planet. The smoothing length $\epsilon$ is $0.6\,r_H$, where $r_H$ is
the Hill radius of the planet. The term $f(s)$ is our filter used to
exclude the neighborhood of the planet given by\,:
\begin{equation}
f(s) = \left[ \exp\left(-10\left(\frac{s}{0.6\,r_H}-1\right)\right) +
1 \right]^{-1}\ .
\label{eq:f}
\end{equation}
It is a smooth increasing function from 0 at $s=0$ to $1$ when
$s\to\infty$, through $0.1$ when $s=0.47\,r_H$, $1/2$ when
$s=0.6\,r_H$, and $0.9$ when $s=0.73\,r_H$ \citep[see Figure~2
of][]{Crida-etal-2008}.

\subsection{Disk and planets parameters}
\label{subsec:disk-pl}

The Solar Nebula found by \citet{Desch2007} has the following
properties. Assuming that Jupiter formed at $5.45$~AU, Saturn at
$8.18$~AU, Uranus at $14.2$~AU and Neptune at $11.5$~AU, the gas
density should be\,:
\begin{eqnarray}
\label{eq:Sigma_0}
\hspace{-1.5cm}\Sigma_0(r) & = & 343(r/10 {\rm
AU})^{-2.168}\ {\rm g.cm}^{-2}
\end{eqnarray}
$$ = \  5.68\times10^{-3}(r/1 {\rm
AU})^{-2.168}\ M_\odot {\rm AU}^{-2}\ .$$

This is about 10 times more dense than the \citet{Hayashi1981} nebula
at 5~AU, and 6.4 times more dense at 10~AU, and 3 times more dense at
30~AU. The steep density slope makes the disk a decretion disk,
viscously spreading outwards, fed by the internal parts. The steady
state profile found by \citet{Desch2007} has a slightly different
shape, but we use the power law for convenience. The two profiles are
almost identical in the giant planets region. The outer edge of the
disk is found to be located at $61$~AU from the star.

The temperature is $150(r/1 {\rm AU})^{-0.429}\,K$, which corresponds
to an aspect ratio of
\begin{equation}
(H/r)_0=0.05(r/1{\rm AU})^{0.2855}\ .
\label{eq:H/r_0}
\end{equation}
The viscosity is given by an
$\alpha$ prescription \citep{Shakura-Sunyaev-1973}, with $\alpha =
4\times 10^{-4}$.

With these characteristics, the disk profile remains almost unchanged
for nearly 10 millions years, which leaves time for {the solid
cores of the planets to reach their isolation masses and accrete their
atmosphere}. However, this also leaves {the planets} time to
migrate.

In our simulations, the planets are initially located on circular
orbits at the position from which the \citet{Desch2007} disk is
derived (see above). The planets masses are grown smoothly from 0 to
their present masses over $30$ years at the beginning of the
simulation. The planets are not accreting gas from the disk.

During the first 100 orbits of Jupiter (1274 years), the planets do
not feel the disk potential, and therefore don't migrate. During this
time, the planets launch a wake, open a gap, perturb the disk. {
This is for the disk to adapt to the planets potential and reach an
equilibrium state. Then, the planets} are released under the influence
of the disk, and start their migration.

\subsection{Expected migration}

The aspect ratio is $8.1\%$ at the location of Jupiter, and $9.1\%$ at
the location of Saturn. The Reynolds number
$\mathcal{R}=r^2\Omega/\nu$ is $3.8\times10^5$ at the location of
Jupiter, and $3.02\times10^5$ at the location of Saturn. Denoting
$\mathcal{P}=\frac{3}{4}\frac{H}{r_H}+\frac{50}{q\mathcal{R}}$~, we
find $\mathcal{P}=0.876+0.131\approx 1$ for Jupiter, and
$\mathcal{P}=1.498+0.583\approx 2$ for Saturn. The unified criterion
to open a gap of depth $90\%$ of the unperturbed density is
$\mathcal{P}\lesssim 1$ \citep{Crida-etal-2006}. Therefore, we expect
Jupiter to open a non-empty gap, and Saturn not to perturb the density
profile significantly.

Consequently the mechanism used by \citet{MC07} to prevent Jupiter and
Saturn migration should not work\,: it requires that the two planets
lie in a wide common gap in mean motion resonance. On the contrary,
Saturn should be here in type~I migration, like Uranus and
Neptune. Given that these three planets are massive ($15$ to almost
$100$ Earth masses), and that the disk density is high, their type~I
migration should be fast.

The isothermal horseshoe drag exerts on the planet a torque
proportional to the logarithmic gradient of the vortensity\,:
$\left({\rm d}\ln (\Sigma/B)\right)\,/\,\left({\rm d} \ln r\right)$,
where $B$ is the second Oort constant \citep{Ward1991,Masset2001}. In
Keplerian rotation, $B=\Omega/4$. Thus, in the \citet{Desch2007}
nebula, $\Sigma/B\propto r^{-0.668}$, and ${\rm d}\ln (\Sigma/B)/{\rm
d} \ln r=-0.668$. Due to the steep negative slope of the density
profile, the corotation torque is negative, while it is zero if
$\Sigma\propto r^{-1.5}$ and positive if the slope is shallower. Such
a disk density slope should therefore enhance the type~I inwards
migration of the planets with respect to more classical disks, in
particular the \citet{Hayashi1981} MMSN.

The high mass of the disk and the fact that Jupiter opens a small gap,
makes it likely that Jupiter falls in the type~III, runaway migration
regime \citep{MassetPapaloizou2003}. Indeed, the mass of gas present
in the coorbital region of Jupiter in the unperturbed disk is about 4
times the mass of Jupiter\,; therefore, the coorbital mass deficit may
exceed the mass of the planet, which is the condition for the runaway
migration.

In summary, in the Nebula found by \citet{Desch2007}, the four giant
planets of the Solar System should migrate inwards on a short
timescale. To check this, and to study the possible interactions
between the four planets in the nebula, we have performed numerical
simulations, presented in the next sections.

\section{Models with locally isothermal equation of state}
\label{sec:isoth}

In this subsection, we use a locally isothermal equation of state
$P={c_s}^2\Sigma$, with $c_s$ the sound speed. The sound speed is a
function of the distance to the star given by
${c_s}^2=(H/r)^2GM_\odot/r\propto T$, which is not evolving with
time. The energy equation is not computed. The 2D grid ranges between
$1.6$ and $23.5$~AU, with a resolution of $\delta r/r = 0.01$, so that
it is divided in 270 rings, and 628 sectors.

\begin{figure}
\plotone{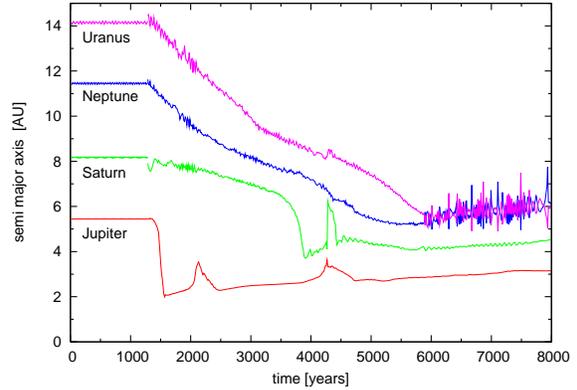}
\caption{Migration path of Jupiter, Saturn, Uranus and Neptune in the
\citet{Desch2007} nebula, with locally isothermal EOS and resolution
of $10^{-2}$.}
\label{fig:isoth_lowres}
\end{figure}

The planets are set according to section~\ref{subsec:disk-pl}. The
migration path obtained is displayed in
Figure~\ref{fig:isoth_lowres}. As soon as released, Jupiter enters a
type~III migration regime and reaches $2$~AU in $300$ years. There, it
stops due to interaction with the boundary of the 2D grid. If the grid
had extended further inwards, there is no reason why Jupiter would not
have reached the Sun.

Saturn also migrates inwards, at first in type~I migration because it
hardly perturbs the density profile. It enters the type~III migration
regime only at $t=3600$ years. Indeed, at this time, Saturn has
reached $6.2$~AU, where $\mathcal{P}_{\rm Saturn}=1.88$. Then, the dip
dug by Saturn in the disk profile is deeper than before, the {\it
coorbital mass deficit} can reach the mass of Saturn, and the runaway
process can start.

When Saturn reaches $3.7$~AU, at $t=3900$ years, it is caught in the
5:3 mean motion resonance with Jupiter\,: for 300 years, the ratio
between their semi major axes is $a_S/a_J=(5/3)^{2/3}$ (where $a$
denote the semi major axis, and the subscript refers to the name of
the planet). This breaks the runaway. At $t=4260$ years, Saturn has a
close encounter with Jupiter and is kicked out. Then, it starts again
an inwards migration until it is blocked again by Jupiter\,: from
$5600$ on, the ratio between their semi major axes is almost constant
$a_S/a_J \approx 1.43$. This may be due to a mean motion resonance or
to indirect interactions, Jupiter perturbing the disk and thus the
motion of Saturn\,; in particular, the corotation torque should be
positive and strong on the outer edge of the gap of Jupiter, which
repels Saturn \citep[see][]{Masset-trap-2006}. The two planets then
start to migrate slightly outwards in a common gap, like in
\citet{MS01}.

Neptune and Uranus are in the type~I migration regime. Between $2600$
and $3600$ years, Neptune is slowed down by Saturn\,: the ratio of the
semi-major axis of the two planets remains constant equal to
$a_N/a_S=1.17\approx (5/4)^{2/3}$ during this period. Between $3060$
and $3520$, Uranus is also slowed down, with $a_U/a_S=1.39\approx
(5/3)^{2/3}$, and $a_U/a_N\approx (4/3)^{2/3}$. This shows likely
captures of Neptune in the 5:4 resonance with Saturn, and Uranus in
the 5:3 with Saturn or the 4:3 with Neptune. However, a mean motion
resonance can not be maintained if the involved bodies move away from
each other. When Saturn accelerates its inwards migration, Neptune and
Uranus stay behind.

Once Saturn has moved inward, the two ice giants accelerate again
their inwards migration. Neptune goes faster and reaches an other
resonance with Saturn. From $4800$ years on, their semi-major axis
ratio is constant equal to $(7/5)^{2/3}$.

At time $5900$ years, Uranus reaches Neptune. {The two planets
then share the same average semi-major axis, equal to
$1.5^{2/3}\,a_S$, which is characteristic of a 3:2 mean motion
resonance with Saturn. Their configuration will be analyzed in more
detail in section~\ref{sub:UN}.}

\subsection{High resolution}
\label{subsec:HR}

Because some authors \citep{DABL05} have found that type~III migration
is resolution dependent, we have performed the same experiments with
a better resolution of $\delta r/r = 10^{-2.5}$. The 2D-grid is
also extended inwards and ranges now from $r=1$~AU to $r=23.5$~AU. The
2D grid is now divided in $1000$ rings and $1987$ sectors. The 1D grid
is divided in $1811$ rings between $0.2$ and $61$~AU.

\begin{figure}
\plotone{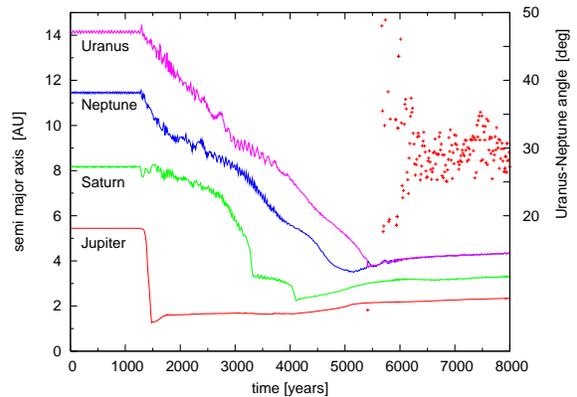}
\caption{Migration path of Jupiter, Saturn, Uranus and Neptune in the
\citet{Desch2007} nebula, with locally isothermal EOS and resolution
of $10^{-2.5}$ (curves, left $y$ axis). The crosses show the angle
between Uranus and Neptune when they share the same orbit (right
$y$-axis).}
\label{fig:isoth_bw}
\end{figure}

The result is displayed in Figure~\ref{fig:isoth_bw}. It is very similar
to the previous case. Jupiter still migrates in type~III migration as
soon as released, and Saturn also undergoes a runaway migration
episode, but a few hundreds of years earlier than before. The planets
go further inwards due to the extended 2D grid\,; this shows that
their stop is a numerical artifact, as expected. The global result is
not affected by the resolution. In fact, Figures~\ref{fig:isoth_lowres}
and \ref{fig:isoth_bw} look similar, and the same explanations hold.

From $4100$ years on, Saturn and Jupiter migrate slightly outwards in
the 5:3 mean motion resonance, with $a_S/a_J=(5/3)^{2/3}$.

Neptune and Uranus have the same migration rate in both resolution
cases, before $3000$ years. After $3000$ years, they migrate faster in
the high resolution case because Saturn is not present anymore to slow
them down. At $t=5400$, a close encounter happens between Uranus and
Neptune (the distance between the two planets is smaller than 0.4
AU). Then, the two planets share the same semi major axis {
again. More detail on their configuration will be provided in
section~\ref{sub:UN}.}

After $10^4$ years, the four planets have escaped a fall into the Sun,
thanks to the inner edge of the 2D grid. They have reached a fully
resonant configuration in which their inwards migration is
prevented. This is roughly similar to
\citet{Morby-etal-2007}. However, here the four giant planets lie
within 7~AU from the Sun, with Jupiter being at $3.6$~AU. This is
incompatible with the present structure of the main asteroid belt. In
addition, the last planet is not stopped by a resonance with Saturn or
with the last but one planet, so that the two ice giants share the
same orbit. This is because the disk is more massive here than in
\citet{Morby-etal-2007}, and the gaseous torques are stronger than the
more distant resonances. It is very unlikely that this final
configuration may lead to a late global instability that would drive
the planets on their present orbits.

\subsection{Getting rid of the hot Jupiter}

In the previous simulations, the migration of Jupiter was stopped
because the giant planet reached the inner edge of the 2D
grid. Consequently, what happens after this event is admittedly
interesting, but somehow artificial. In a new simulation, the planets
are removed when they reach $a<2$~AU.

\begin{figure}
\plotone{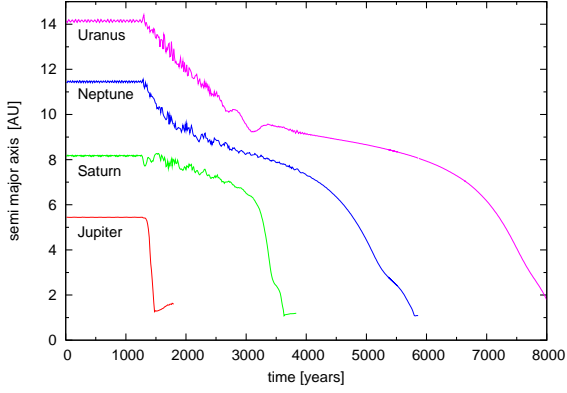}
\caption{Migration path of Jupiter, Saturn, Uranus and Neptune in the
\citet{Desch2007} nebula, with locally isothermal EOS and resolution of
$10^{-2.5}$. The planets are removed from the simulation when they reach
the 2D-grid inner edge.}
\label{fig:isoth_bw_apu}
\end{figure}

The results are displayed in Figure~\ref{fig:isoth_bw_apu} for the
higher of two resolutions used above ($\delta r/r=10^{-2.5}$). Jupiter
being not present anymore, Saturn goes on migrating inwards and
finally reaches $a_S<2$~AU as well. Then, Uranus and Neptune migrate
inwards freely. The differential Lindblad torque, responsible for
type~I migration, is proportional to $\Sigma r^4\Omega^2(H/r)^{-2}$\,;
thus it increases as $r^{-1.74}$ when the planets approach the
Sun. This explains their accelerating migration. Within $8000$ years,
they have migrated inwards all the way to the grid edge too. The same
happens at lower resolution.

With this more realistic way of dealing with the planets at the inner
edge of the 2D grid, the four giant planets are lost in less than
$10^4$ years.

\section{Non isothermal disk}
\label{sec:energ}

It has been shown recently that computing the energy equation can
change dramatically the corotation torque, and thus the migration rate
of low mass planets
\citep{Paardekooper-Mellema-2006,Baruteau-Masset-2008}. Planets in
type~II migration should not be perturbed much, but this process may
be critical for Neptune mass planets \citep{Kley-Crida-2008}. In the
case considered in this paper, we expect this effect to be also
significant for Saturn and Jupiter because their gap is not empty. To
check this, we have implemented in FARGO-2D1D the computation of the
Energy equation as done in the FARGO-ADSG version.

The Energy equation is the following, with $e$ the surface density of
internal energy of the gas, $\vec{v}$ the velocity vector, and $P$ the
pressure\,:
\begin{equation}
\frac{\partial e}{\partial t} + \nabla(e\vec{v}) = 
-P\nabla\vec{v} + Q_+ + Q_-\ ,
\label{eq:E}
\end{equation}
where $Q_+$ and $Q_-$ are heating and cooling terms, respectively.

The heating term comes from the viscous heating. The cooling term is
given by a vertical black body emission\,: $Q_-=2\sigma_R T^4 /
\kappa\Sigma$, where $\sigma_R$ is the Stefan-Boltzmann constant, $T$
is the temperature, and $\kappa$ is the opacity. The opacity is chosen
such that the temperature profile given by \citet{Desch2007} is an
equilibrium profile of the unperturbed disk (in which $Q_+ =
9\Sigma_0\nu\Omega^2/4$)\,:
\begin{equation}
\kappa = \frac{8}{9}\frac{{\sigma_R(H/r)_0}^8}{\nu\Sigma^2 r}
\label{eq:kappa}
\end{equation}
with $(H/r)_0$ given by Eq.~(\ref{eq:H/r_0}).

The cooling time $\tau_{\rm cool}=e/Q_-$ is then
$(10/9\alpha)\,\Omega^{-1}$, $\tau_{\rm cool}=2.78\times
10^{3}\,\Omega^{-1}$. For Uranus and Neptune, the horseshoe libration
time $\tau_{\rm lib}=8\pi r_p/(3\Omega x_s)$ is $326\,\Omega^{-1}$ at
12~AU, with $x_s=1.16r_p\sqrt{q/(H/r)}$
\citep[][Eq.~(3)]{Masset-etal-2006}. The cooling time being large with
respect to the libration time, the thermal effect on the corotation
torque should saturate quickly and the migration of the ice giants
should not proceed outwards like in \citet{Kley-Crida-2008}.

Our equation of state is, with $\gamma$ the adiabatic index ($\gamma =
1.4$)\,:
\begin{equation}
P = {c_s}^2 \Sigma / \gamma = (\gamma - 1) e\ .
\label{eq:EOS}
\end{equation}
Now, the sound speed $c_s$ is not fixed but determined at every
time-step by the local temperature, which is evolving with time.

\begin{figure}
\plotone{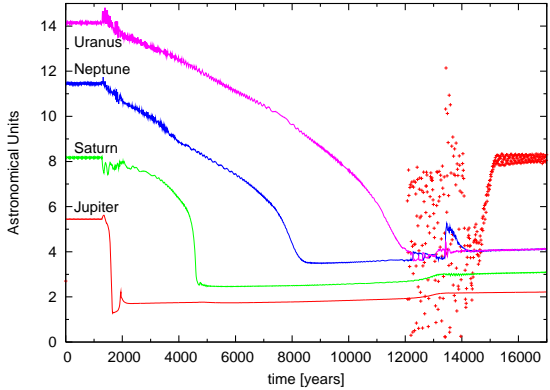}
\caption{Migration path of Jupiter, Saturn, Uranus and Neptune in the
\citet{Desch2007} nebula, with Energy equation computed, and resolution
of $10^{-2}$ (curves). Crosses\,: Uranus-Neptune distance.}
\label{fig:2_lowres}
\end{figure}

The same experiments as in previous section are computed, with this
energy equation and equation of state. The result at lower resolution
is shown in Figure~\ref{fig:2_lowres}. Jupiter and Saturn still
undergo a type~III migration episode. The migration of Uranus and
Neptune is slower than with a locally isothermal disk, but is still
directed inwards. In the end, the situation is basically unchanged
with respect to previous section\,: Jupiter is blocked by the edge of
the 2D grid, Saturn is blocked by Jupiter, Uranus and Neptune share a
common orbit in resonance with Saturn.

\begin{figure}
\plotone{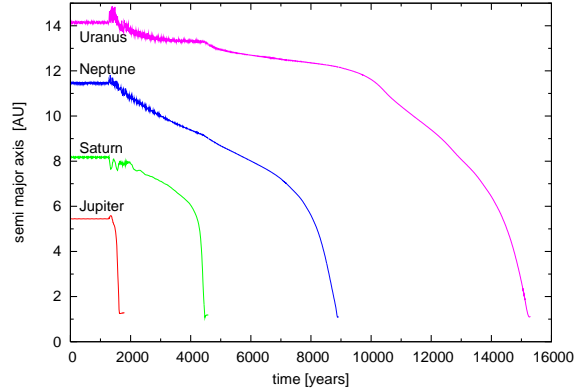}
\caption{Migration path of Jupiter, Saturn, Uranus and Neptune in the
\citet{Desch2007} nebula, with Energy equation computed, and
resolution of $10^{-2.5}$. The planets are removed when they encounter
the inner edge of the 2D grid.}
\label{fig:2_highres_apu}
\end{figure}

With a higher resolution ($10^{-2.5}$), and removing the planets when
they reach the inner edge of the 2D grid, one gets the migration paths
displayed in Figure~\ref{fig:2_highres_apu}. Jupiter and Saturn
disappear below $1.3$~AU within $4500$ years. The migration of Uranus
and Neptune is a bit slower than in the previous case. However, as
soon as Saturn starts its type~III inwards migration, the migration of
the ice giants accelerates. Then the migration of Neptune is similar
to the previous case (with a 800 years delay), while that of Uranus is
much slower ($\dot{a}_U\approx 2\times 10^{-4}$ AU.year$^{-1}$). After
$8800$ years, Neptune reaches $1.1$~AU and is declared lost. Uranus goes
on migrating slowly inwards, accelerates after Neptune is suppressed,
{and reaches $1.3$~AU after $15200$ years}.

\section{Interactions between the planets}

\subsection{Jupiter and Saturn}
\label{sec:MMR}

Previous sections have shown that if the 4 giant planet of the Solar
System are placed in the \citet{Desch2007} nebula in the starting
configuration of the Nice model and released simultaneously, they
migrate inwards and disappear in the Sun in about $ 10^4$ years. {
In this section, we try to prevent this by releasing the planets in
sequence, and varying the disk aspect ratio. We aim at reproducing the
best candidate mechanism so far to prevent the inwards migration of
Jupiter and Saturn\,:} \citet{MS01} and \citet{MC07} have shown that
when Jupiter and Saturn orbit in a mean motion resonance in a common
gap, they can avoid type~II migration or even migrate outwards.

Therefore, we have performed simulations in which Jupiter is held on a
fixed circular orbit longer than Saturn and the ice giants. The three
lighter planets migrate inwards. Once Saturn's semi major axis has
reached a constant value in terms of Jupiter semi major axis (which
betrays a resonance), we release Jupiter as well. Immediately, Jupiter
migrates inwards in type~III migration and disappears\,; and the three
other planets migrate inwards again. This is because the considered
disk is too thick to enable Jupiter to open a deep and wide gap,
embracing also Saturn.

So, we decrease the aspect ratio to $H/r=0.04$. As the viscosity is
given by an $\alpha$-prescription, the viscosity is also smaller. In
fact, $\mathcal{P}_{\rm Jupiter}=0.467$. The planets are maintained on
a fixed orbit for 5250 years (500 Jupiter orbits). At this time,
Jupiter has opened a deep and wide gap, the density at the bottom of
which is a bit less than $2\%$ of the unperturbed density. Saturn has
opened a partial gap, of depth a half of the unperturbed density, and
the two gaps have merged (see the solid profile in
Figure~\ref{fig:profile_AR0.04}, compared to the dot-dashed initial
profile). This kind of gap suits the \citet{MS01} mechanism\,: Jupiter
and Saturn lie together between the inner and the outer disk.

First, only Saturn is released. One expects than Saturn, repelled by
the outer disk, would migrate inwards until it encounters a mean
motion resonance with Jupiter. However, Saturn goes immediately in
type III migration outwards and reaches $10$~AU. Twenty orbits later,
Jupiter, Uranus and Neptune are released as well. Jupiter runs away
inwards and reaches $2$~AU in about a hundred years. If the 4 planets
are released simultaneously after 500 orbits of Jupiter, the same
thing happens.

In an other attempt, Uranus and Neptune are not considered, and the
aspect ratio is set constant equal to $0.05$, like in the almost
stationary solution found by \citet{MC07}. Saturn is placed on a
circular orbit at $10$~AU, and let free to migrate after 2550 years
(200 Jupiter orbits), while Jupiter is still held on a fixed orbit at
$5.45$~AU. Saturn migrates inwards in type~III migration, and then is
blocked by Jupiter. Then, we release Jupiter as well after 5100 years
(400 Jupiter orbits). Again, Jupiter runs in type~III migration into
the Sun as soon as released.

{Decreasing even more the viscosity and aspect ratio of the disk
may help, but $H/r=0.04$ and $\alpha=4\times 10^{-4}$ are already
rather small compared to the standard values inferred from
observations.} In fact, there are two reasons why the \citet{MS01}
mechanism can not work in the considered disk, even if one assumes
that it is cool enough to enable Jupiter to open a wide gap\,:
\begin{enumerate}
\item The density of this disk is so high that Jupiter (or Saturn) can
easily enter the runaway type~III migration regime. In that case,
Jupiter is driven inwards faster than Saturn, and the putative
resonance is broken.
\item This disk is a decretion disk. The \citet{MS01} mechanism
prevents a pair of planets from inwards type~II migration. It works
when the innermost planet is more massive than the outer one, and
repels the later outwards against the accreting outer disk. Here, the
global evolution of the disk is directed outwards. A more massive
Saturn than Jupiter would be needed to prevent the outwards type~II
migration of the two planets.
\end{enumerate}

\subsection{Uranus and Neptune}
\label{sub:UN}

{

In the above simulations, when the migration of Jupiter is
artificially stopped at the edge of the 2D grid, Uranus and Neptune
always end on the same orbit, captured in the same mean motion
resonance with Saturn. This strange configuration is worth a more
detailed analysis.

In the first case presented (locally isothermal disk, with resolution
$\delta r/r = 10^{-2}$), the distance between the two planets
--\,shown as crosses in Figure~\ref{fig:UN}\,-- is bounded by $0.1$~AU
after $t=6000$ years.} The Hill radius of the planets is
$r_H=a_p(M_p/3)^{1/3}=0.14$~AU at $a_p=5.5$~AU. This indicates a
satellite motion of the planets. This kind of configuration requires a
lot of damping because the planets were of course unbound
gravitationally at the beginning. The damping is here provided by the
disk, and is visible between $5900$ and $6000$ years in
Figure~\ref{fig:UN}\,: the maximum distance between the planets
decreases until it becomes smaller than the Hill radius. The high
density of this nebula makes such a satellite capture possible.

{In the second case (locally isothermal, $\delta r/r =
10^{-2.5}$), the planets are not satellites of each other. Here,} the
distance between Uranus and Neptune is slightly oscillating around
$2$~AU after $6000$ years. They share the same orbit, with an
eccentricity smaller than $0.04$ and Uranus leading about $30^\circ$
before Neptune\,: {the angle between Uranus and Neptune, seen from
the Sun, is plotted as crosses in Figure~\ref{fig:isoth_bw}, with
respect to} the right $y$-axis. The two planets are not at a Lagrange
point of the other one. This unusual configuration may be due to the
mean motion resonance with Saturn, or more probably to the strong
dissipation by the disk, that modifies the local dynamics. Once the
disk will have evaporated, these two planets should have a tadpole
motion around one of their $L_4/L_5$ Lagrange points. Such a
configuration have already been studied, for instance by
\citet{Thommes2005,Beauge-etal-2007,Cresswell-Nelson-2008}. Generally,
it is assumed that a giant planet captures a migrating terrestrial
planet, that may then grow. In contrast, here the effect of the giant
planet is to put two 15 Earth mass planets together on the same
orbit. This is a new way of forming Trojan planets.

{In the third case (with energy equation, and $\delta r/r =
10^{-2}$),} the distance between the two ice giants (shown as the $+$
symbols in Figure~\ref{fig:2_lowres}) settles to about 2 times their
semi-major axis. This indicates that the two planets are in
opposition, librating around their $L_3$ point. Once again, this
configuration is not possible in the absence of gas dissipation or of
the other planets, because the $L_3$ point is unstable. If the other
planets and the disk disappear, these two planets should have
coorbital horseshoe orbits.

{These three original outcomes open new possibilities for
extra-solar systems.  They do not fit the Solar System configuration,
though. Whether or not a scenario like the Nice model may apply from
such 1:1 resonances, is not clear. In particular, the configurations
with Uranus and Neptune at opposition or at 30 degrees on the same
orbit are probably unstable on the short term in absence of the gas
disk. This is thus unlikely to yield to a {\it late} instability, as
required. In the case where the two planets are satellite of each
other at a distance $d=0.1$~AU, the energy required to separate them,
$(GM_{\rm U}M_{\rm N}/d-GM_{\rm U}M_{\rm N}/r_H)/2$, is roughly
equivalent to the energy required to eject a Moon size body out of the
Solar System from a 10 AU orbit. Therefore, it is likely that
they stay in this satellite configuration.  }

\section{Migration in the \citet{Weiden77} - \citet{Hayashi1981} nebula}
\label{sec:Hayashi}

\begin{figure}
\plotone{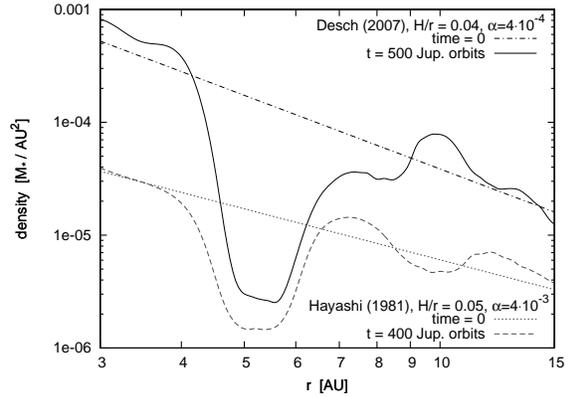}
\caption{Top black lines\,: Initial density profile (dot-dashed,
straight line) and profile after 500 Jupiter orbits (solid curve) in
the \citet{Desch2007} nebula with $H/r=0.04$ (see
section~\ref{sec:MMR}). Bottom gray lines\,: Initial density profile
(dotted, straight line) and profile after 400 Jupiter orbits (dashed
curve) in the \citet{Hayashi1981} nebula (see
section~\ref{sec:Hayashi}).}
\label{fig:profile_AR0.04}
\end{figure}

\begin{figure}
\plotone{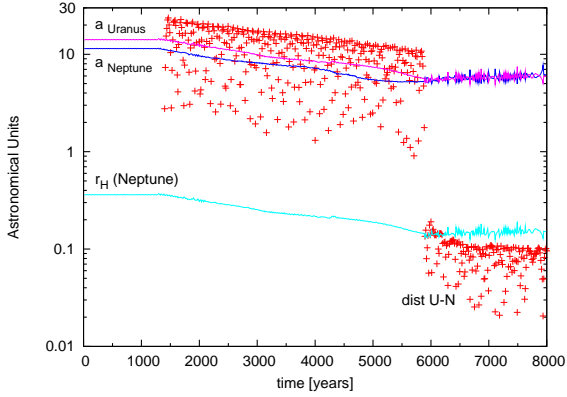}
\caption{Migration path of Uranus and Neptune taken from
Figure~\ref{fig:isoth_lowres} (two top curves labeled ``a$_{\rm
Uranus}$'' and ``a$_{\rm Neptune}$'' respectively). The size of
Neptune Hill radius is the bottom light blue curve (``$r_{\rm H}$
(Neptune)'' ). The distance between the two planets is shown with $+$
symbols, labeled ``dist~U-N'').}
\label{fig:UN}
\end{figure}

For comparison, the same experiment as in section~\ref{sec:isoth} has
been performed in the {standard Minimum Mass Solar Nebula. The initial
density is the one quoted in introduction, that is\,:
\begin{equation}
\Sigma(r)=1.9125\times 10^{-4} (r/1\,{\rm AU})^{-3/2}\ M_\odot.{\rm
AU}^{-2}\ .
\label{eq:Hayashi}
\end{equation}
This profile appears in Figure~\ref{fig:profile_AR0.04} as the
straight, gray, dotted line.}

The equation of state is locally isothermal. {The aspect ratio is
given by Eq.~(\ref{eq:H/r_0}), and the viscosity is given by an
$\alpha$-prescription with $\alpha=4\times 10^{-4}$.} The planets are
initially placed on circular orbits at their present position\,:
$a_J=5.2$~AU, $a_S=9.6$~AU, $a_U=19.2$~AU, $a_N=30$~AU. The resolution
is $10^{-2}$, the 2D grid extends radially from $2.15$ to $52$~AU,
while the 1D grid covers the range $[0.312\,;\,104]$~AU.

\begin{figure}
\plotone{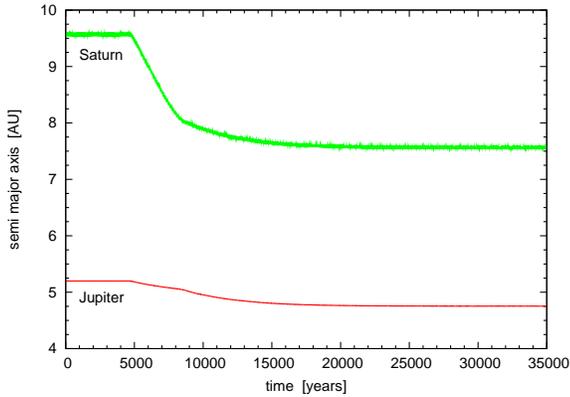}
\caption{Migration path of Jupiter (bottom red curve) and Saturn
 (top green curve) in {a disk with the \citet{Hayashi1981} density
 profile}, $H/r=0.05(r/1{\rm AU})^{0.2855}$ and $\alpha=4\times
 10^{-4}$ like in the \citet{Desch2007} nebula. The Equation of State
 is locally isothermal. The resolution is $\delta r/r=10^{-2}$.}
\label{fig:Hayashi_D}
\end{figure}

The planets do not feel the disk potential during 400 Jupiter orbits
(4740 years), and are then released. {The evolution of the semi
major axes of Jupiter and Saturn are displayed in
Figure~\ref{fig:Hayashi_D}. In spite of the fact that the shape of the
gaps opened by the planets are similar to the ones in the
\citet{Desch2007} disc \citep[the width and depth depend only on
$M_p$, $H/r$ and $\nu$, see][]{Crida-etal-2006}, no type~III migration
take place, because this disc is not massive enough. Saturn migrates
faster than Jupiter\,; about 2600 years after the release, the two
planets enter in 2:1 mean motion resonance, which slows Saturn
down. After $25000$ years, the migration stops at $a_{\rm J}=4.75$,
$a_{\rm S}=7.57$~AU.  }

{In an other, longer term simulation, we change the aspect ratio
to $H/r=0.05$ and the viscosity to $\alpha = 4\times 10^{-3}$. The low
aspect ratio enables Jupiter to open a significant gap, and Saturn a
shallow one. The density profile after 400 Jupiter orbits can be seen
as the bottom, gray, dashed line in Figure~\ref{fig:profile_AR0.04}.}
The migration paths are displayed in Figure~\ref{fig:Hayashi} {for
$10^5$ years}. Jupiter migrates inwards in type~II migration, while
Saturn migrates faster. After 6400 years, Saturn encounters the 2:1
mean motion resonance with Jupiter, but is not caught in it. At about
$10^4$ years, Saturn encounters the 5:3 mean motion resonance with
Jupiter\,; from this time on, $a_S/a_J\approx (5/3)^{2/3}$. The gaps
of the two planets merge. Then, the \citet{MS01} mechanism takes
place, and the two planets have {very little} migration, like in
\citet{MC07}\,: their {semi major axes don't change by more than
1~AU in $10^{5}$~years.}

Note that the stop of the migration of Jupiter at $4.5$ {(or
$4.75$)}~AU is not due to the edge of the 2D grid\,: the radius of the
inner edge of the 2D grid is about a half of the semi major axis of
Jupiter. The presence of Saturn is responsible for the salvation of
Jupiter. To check this, the same simulation has been performed with
Jupiter alone, without the other three planets. The migration path is
displayed in Figure~\ref{fig:Hayashi} as the orange dashed line. In
this case, Jupiter has an unperturbed type~II migration, and {
reaches 3 AU in $40\,000$ years, $2.7$ AU in $50\,000$ years.}

For comparison, the migration path of Jupiter in
Figure~\ref{fig:2_highres_apu} is displayed {also in
Figure~\ref{fig:Hayashi}} as a dotted red line. It
looks like a free fall towards the Sun.

Uranus and Neptune are migrating inwards in type~I migration. Their
migration speed is slower than in the previous cases, because the
density is moderate {and the density slope is such that the
isothermal corotation torque is zero. When approaching Saturn, Uranus
slows down and almost stops between 9 and 10 AU. Then, Neptune is
caught in the 2:3 mean motion resonance with Uranus at $8.5\times
10^4$ years\,: from this date on, their semi-major axis ratio is
$1.5^{2/3}$.}

\subsection{Conclusion}

Migration also occurs in the \citet{Hayashi1981} nebula, and it is not
possible that the giant planets stay where they {form} until the
present time. Therefore, the \citet{Hayashi1981} nebula is not more
consistent with planetary migration than the \citet{Desch2007}
one. However, no runaway migration takes place, and the Jupiter-Saturn
pair {can} avoid migration all the way to the Sun {for some
reasonable disk parameters}. Then, a scenario like the one presented
in \citet{Morby-etal-2007} is possible. In this article, the authors
find 6 fully resonant configurations {of the four giant planets}
in which the migration is prevented. After disk dissipation, two of
these configurations can lead to a late global instability of the
planetary system, compatible with the Nice model.

\section{Discussion}
\label{sec:discussion}

\subsection{Energy Equation and cooling law}

{To slow down migration in the \citet{Desch2007} nebula, one can
think of taking} the energy equation into account with an other
opacity law. This would change the thermal structure of the disk. With
a shorter cooling time, the temperature and the aspect ratio
decrease. This has basically two effects\,: (i) Saturn and Jupiter
would open a deeper gap, but our locally isothermal simulations with
$H/r\leqslant 0.05$ show that it may not be enough to prevent their
migration\,; (ii) the thermal horseshoe effect would not saturate if
$\tau_{\rm cool} \sim \tau_{\rm lib}$, which could lead to a positive
corotation torque on the planets, particularly Uranus and Neptune.

This should be the subject of a future, dedicated study. However, it
is rather unlikely that the four planets stay exactly on their initial
orbit until the disk dissipates, in order for the Nice model to take
place.

\subsection{Accretion}

{

In all the simulations presented so far, the planets were not allowed
to accrete. This may appear unrealistic at first sight, but on the
other hand, accretion must have stopped anyhow before the masses of
the planets exceed the present masses. This is certainly a critical
issue for the formation of Jupiter and Saturn, because the accretion
timescale for so massive planets should be very short. In both density
profiles studied here, free accretion of the planets leads to too high
masses. This open problem requires a complete, dedicated study, that
is beyond the scope of this paper. We make the assumption that some
mechanism prevented accretion to go on when Jupiter and Saturn reached
their actual masses. However, a few comments are in order.

In the simulation presented in Figure~\ref{fig:Hayashi}, the gas
density at the bottom of the gap created by Saturn is of the order of
3 times the one at the bottom of the gap of Jupiter. If this situation
lasts a few hundred thousand years, Saturn may accrete more gas than
Jupiter on the long term. This would be a problem for the \citet{MS01}
mechanism\,: a lighter outer planet than the inner one is required for
the migration to be slowed down, stopped, or reversed. In the Solar
System though, we know that the mass of Saturn somehow remained
smaller than a third of the mass of Jupiter. This mass ratio allows
the migration of the pair to be negligible over the disc life time,
with suitable disk parameters. Therefore, the \citet{MS01} mechanism
applies.

An other issue with accretion concerns type~III
migration. \citet{DAL08} have observed that allowing the planet to
accrete, may prevent the runaway migration. However, this is most
likely because their accreting planets grow too fast from low mass
planets in type~I migration to high mass planets in type~II migration
in a clean gap. With accretion turned on, but the planet mass
artificially kept constant to $3\times 10^{-4}M_*$, they find that
type~III migration occurs, which shows that accretion should not
perturb too much the corotation torques responsible for the
runaway. To check this, we have restarted the simulation presented in
section~\ref{subsec:HR} and Figure~\ref{fig:isoth_bw} at the time
where the planets are released, turning accretion by Jupiter and
Saturn on. Accretion is computed using the recipe by \citet{Kley99},
with a timescale to accrete all the gas within $0.45r_H$ of the planet
equal to $1.6$~year, that is $1/8$ Jupiter orbit. The evolution of the
mass and semi major axis of Jupiter are displayed in
Figure~\ref{fig:accr} as the curves with $+$ symbols, compared to the
solid line taken from Figure~\ref{fig:isoth_bw}. Type~III migration is
confirmed (and even enhanced), even if Jupiter reaches 5 Jupiter
masses in a century.

For all these reasons, and even if a deeper study of the accretion
processes would certainly be valuable, we think that our results are
robust as far as this issue is concerned.

}

\section{Summary and conclusion}
\label{sec:conclu}

In this paper, we have confronted the \citet{Desch2007} Solar Nebula
to the planetary migration issue. This disk has a high gas density, a
high density slope, and this profile remains almost unchanged for 10
million years. Therefore, the question of the planetary migration
should not be eluded. The 2D1D version of FARGO is particularly suited
to such a study. The physical parameters of the disk are fixed, but a
broad range of numerical parameters have been studied (resolution,
energy equation, fate of the planets that reach the inner edge of the
2D grid), as well as the influence of the aspect ratio.

The high density slope gives a negative non thermal horseshoe torque,
that accelerates even more the inwards type~I migration of Uranus and
Neptune. The high density permits type~III migration of massive
planets. In particular, we can not avoid an extremely fast inwards
migration of Jupiter. In spite of several attempts, {it seems to
be impossible} to prevent a loss of all the giant planets in less than
$2\times 10^4$ years, {which is 2 to 3 orders of magnitude shorter
than the disk life time}. Taking into account the energy equation with
a cooling compatible with the assumed temperature structure of the
disk does not change the result.

Preventing the migration of the giant planets requires interactions
between the planets. However, as this disk is a massive, decretion
disk, the \citet{MS01} mechanism to prevent Jupiter and Saturn from
migrating inwards can not be applied. Shortly said, we conclude that
the \citet{Desch2007} nebula is incompatible with our present
knowledge of planetary migration.

So, we are facing a problem. Either this new MMSN makes the planetary
migration in the Solar System an even more critical issue and makes
obsolete the former solutions to prevent migration of the giant
planets \citep{MC07,Morby-etal-2007}, or the existence of planetary
migration makes this MMSN questionable (in particular the apparently
unavoidable type~III migration of Jupiter).

\begin{figure}
\plotone{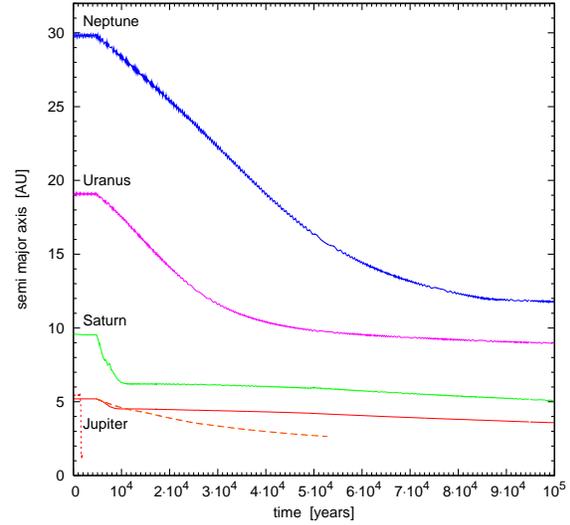}
\caption{Migration path of Jupiter, Saturn, Uranus and Neptune in {
a disk with the \citet{Hayashi1981} density profile}, $H/r=0.05$,
$\alpha=4\times 10^{-3}$. The Equation of State is locally
isothermal. The resolution is $\delta r/r=10^{-2}$. The orange dashed
line corresponds to a simulation where Jupiter alone is migrating in
the {same disk}. The red dotted curve is taken from
Figure~\ref{fig:2_highres_apu}, for comparison.}
\label{fig:Hayashi}
\end{figure}

\begin{figure}
\plotone{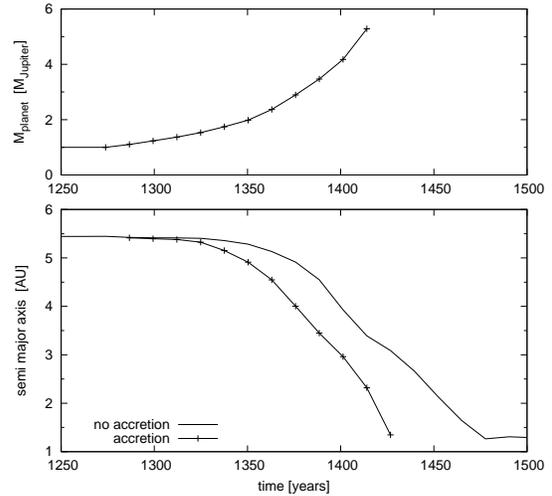}
\caption{Bottom panel\,: Migration path of Jupiter in the
\citet{Desch2007} disk, with locally isothermal equation of state and
resolution $10^{-2.5}$ without accretion (solid line, taken from
Figure~\ref{fig:isoth_bw}), or with accretion with timescale 1.6 year
(solid line with $+$ symbols). Top panel\,: evolution of mass of the
planet in the accreting case.}
\label{fig:accr}
\end{figure}

A possible solution has been laid out in section~\ref{sec:Hayashi}. If
one believes that the Nice model is true, it is clear that the outer
planets should have ended the disk phase in a more compact
configuration than the present one. However, the migration of the
planets can account for a formation on a larger radial range, followed
by a compaction of the configuration of the giant planets, like in
Figure~\ref{fig:Hayashi}. Then, a fully resonant configuration can be
achieved, that can be compatible with the Nice model, like in
\citet{Morby-etal-2007}. In addition, the solid material that built
the giant planets may come not only from the region around their
respective orbits\,: dust drifts inwards in a protoplanetary disk
\citep{Weiden77}, and small bodies migrate as well, so that the giant
planets region may be replenished in solids by the outer parts of the
disk. This enables the formation of Jupiter, Saturn, Uranus and
Neptune in a less massive disk than the Desch nebula, to avoid the
type~III migration of Jupiter. In section~\ref{sec:Hayashi}, we have
seen that this is possible in the old \citet{Hayashi1981}
nebula. However, there is no reason why the giant planets should have
formed exactly where they presently orbit\,; in particular the
presence of an outer cold disk of planetesimals (required in the Nice
model) is problematic if Neptune formed beyond 25~AU. Thus, a new
construction of a MMSN is needed, that takes into account the Nice
model and the planetary formation constraints, like in
\citet{Desch2007}, and also the migration of planets and
planetesimals.

In any case, our results show that planetary migration should be {
considered} in the construction of a MMSN. The location where the
planets form determines the gas density profile of the nebula, which
determines the migration path of the planets, which drives the planets
to a new position after the disk dissipation. This final
configuration, and not the initial one, should be compatible with the
Nice model (or with the present configuration if one does not believe
that the Nice model is true). This idea requires a detailed study,
that is beyond the scope of this paper, but the results presented here
advocate for such a self-consistent construction of the Minimum Mass
Solar Nebula.

\acknowledgments

I wish to thank W. Kley, F. Masset, and A. Morbidelli for
discussions, and W. Kley for reading this manuscript and suggesting a
few improvements. The computations have been performed on the {\tt
hpc-bw} cluster of the Rechenzentrum of the University of
T\"ubingen. B. Bitsch is acknowledged for his help with this
cluster. A. Crida acknowledges the support through the German Research
Foundation (DFG) grant KL 650/7.


\end{document}